\documentclass[preprint,showpacs,preprintnumbers,amsmath,amssymb]{revtex4}

\usepackage{graphicx}
\usepackage{dcolumn}
\usepackage{bm}

\usepackage{feynmf}
\usepackage{slashed}

\usepackage[utf8]{inputenc}

\begin{document}

\title{Dark matter annihilation into right-handed neutrinos and the galactic center gamma-ray excess}

\author{Yi-Lei Tang}
\thanks{tangyilei15@pku.edu.cn}
\affiliation{Center for High Energy Physics, Peking University, Beijing 100871, China}


\author{Shou-hua Zhu}
\thanks{shzhu@pku.edu.cn}
\affiliation{Institute of Theoretical Physics $\&$ State Key Laboratory of Nuclear Physics and Technology, Peking University, Beijing 100871, China}
\affiliation{Collaborative Innovation Center of Quantum Matter, Beijing 100871, China}
\affiliation{Center for High Energy Physics, Peking University, Beijing 100871, China}

\date{\today}

\begin{abstract}

In this paper, we will discuss a specific case that the dark matter particles annihilate into right-handed neutrinos. We calculate the predicted gamma-ray excess from the galactic center and compare our results with the data from the Fermi-LAT. An approximately 10-60 GeV right-handed neutrino with heavier dark matter particle can perfectly explain the observed spectrum. The annihilation cross section $\langle \sigma v \rangle$ falls within the range $0.5$-$4 \times 10^{-26} \text{ cm}^3/\text{s}$, which is roughly compatible with the WIMP annihilation cross section.

\end{abstract}
\pacs{}

\keywords{supersymmetry, vector-like generation, LHC}

\maketitle
\section{Introduction}

The indirect detection experiments of the dark matter (DM) focus on observing the standard model (SM) products from the annihilation or decay of the dark matter particles. Various analyses of the Fermi-LAT data reveal an excess in $1$-$4$ GeV gamma rays from near the center of the milky way \cite{GCE1, GCE2, GCE3, GCE4, GCE5, GCE6, GCE7, GCE8, GCE9, GCE10, GCE11, GCE12}. Fittings to the excess have been carried out (For some examples, see Ref.~\cite{GCE10, VariousChannel, WhereToPut1, WhereToPut2, WhereToPut3, WhereToPut4, WhereToPut5, WhereToPut6, WhereToPut7}) by assuming that the dark matter particles mainly annihilate into $b \overline{b}$, $\tau \overline{\tau}$, $W^+ W^-$, $Z Z$, $h h$, $t \overline{t}$ (For examples, see Ref.~\cite{GCE10, VariousChannel}). Constraints from the dwarf spheroidal (dSph) galaxy candidates are given in Ref.~\cite{dSph1, dSph2, dSph3, dSph4, dSph5, dSph6, dSph7, dsph8}. However, interestingly, there seem to be slight gamma signals from two of the dSphs recently \cite{dSphDiscover1, dSphDiscover2, dSphDiscover3, dSphDiscover4}. Cascade annihilations sometimes appear in the literature in order to avoid the direct detection bounds (For examples, see Ref.~\cite{Cascade1, Cascade2NMSSMN, Cascade3, Cascade4, Cascade5, Cascade6, Cascade7, Cascade8, Cascade9, CascadeAppend1, CascadeRequired1, CascadeRequired2}). Among all these channels, the $b \overline{b}$ channel and the $h h$ offer better-fitted spectrum. The spectrum predicted by the $W^+ W^-$ and the $Z Z$ channels usually peak in relatively higher energy-scales thus being less favoured.

Besides these fully discussed channels, in the literature, models in which the dark matter particles might mainly annihilate into light right-handed neutrinos do exist. For example, the next to minimal supersymmetric standard model (NMSSM) can be extended with right-handed neutrino superfield(s) which only couple(s) with the singlet Higgs \cite{Cascade2NMSSMN, NMSSMN1, NMSSMN2, NMSSMN3, LotsofMistakes, NMSSMN4, NMSSMN5, NMSSMNMy}. Some papers have discussed about the right-handed sneutrino dark matter annihilating into $b \overline{b}$, exotics Higgs pairs. However, there does exist some parameter space that the dark matter particles mainly annihilate into the right-handed neutrinos. Ref.~\cite{BML1, BML2, BML3, BML4} also proposed a model in which dark matter particles mainly annihilate into the right-handed neutrinos through t-channel $Z^{\prime}$ mediators.

In the type I see-saw mechanisms \cite{SeeSaw1, SeeSaw2, SeeSaw3, SeeSaw4, SeeSaw5}, right-handed neutrinos slightly mix with the SM light neutrinos and mainly decay through $h/Z+\nu_{e, \mu, \tau}$, $W^{\pm} + l^{\mp}$ channels by the mixing with the SM neutrinos. If the dark matter particles near the galactic center mainly annihilate into the right-handed neutrinos, these right-handed neutrinos then decay and finally produce the gamma-ray photons detected by the Fermi-LAT. If the mass of the right-handed neutrino $ m_N \gtrsim 100 \text{ GeV}$, it will mainly decay into on-shell $W/Z$ bosons, raising the position of the peak value. In fact, a simple simulation by the micrOMEGAs\cite{micrOMEGAs} shows that in such cases, the predicted gamma spectrum are no better then those in the $W^{+} W^{-}$ and $ZZ$ cases.

If the mass of the right-handed neutrino is less then $m_{W}$, which is just the case to be discussed in the following text, it will decay through off-shell $h/Z/W^{\pm}$ bosons. In order to deal with the off-shell $W$/$Z$/$H$ cases, we use MadGraph5\_aMC@NLO2.3.2\cite{MadGraph} to generate the three-body decay events of one right-handed neutrino at rest in the parton-level, and then input the event file into PYTHIA~8.212\citep{PYTHIA82} to do parton-shower, hadronization, decay process and finally boost the photon spectrum in order to compare the gamma-ray spectrum with the one from Ref.~\cite{GCE12}.

\section{Right-handed neutrino models}

If the dark matter particles $\chi$ with the mass $m_{\chi}$ mainly annihilate into right-handed neutrinos $N$, the key to acquire the gamma-ray spectrum is to calculate the decay processes of the $N$. We input the following Lagrangian into FeynRules 2.3\cite{FeynRules},
\begin{eqnarray}
\mathcal{L} \supset \frac{1}{2} \overline{N} \gamma^{\mu} \partial_{\mu} N - \frac{1}{2} m_N \overline{N} N - ( y_{i} \overline{l}_{Li} \cdot \tilde{H} N + \text{h.c.} ), \label{NLagrangian}
\end{eqnarray}
where $m_N$ is the mass of the right-handed neutrino. $l_{Li}$'s are the left-handed leptonic $SU(2)_L$ doublets. $i$ runs from 1-3 to indicate the $e$, $\mu$, $\tau$ generation. $\tilde{H} = i \tau_2 H^{*}$ is the SM Higgs doublet field, where $\tau_2$ is one of the Pauli matrix. Although in the standard type-I see-saw mechanisms, more than one right-handed neutrinos are needed in order to generate a complete neutrino mass spectrum, in this paper, we assume that only one right-handed neutrino is lighter then the dark matter particle for simplicity. Therefore, it can be produced on-shell during the annihilation processes. If one would like to discuss the cases of more than one right-handed neutrinos, he could just linearly combine the spectrum corresponding with each right-handed neutrino. Thus, Eqn.~(\ref{NLagrangian}) can summarize the features of the right-handed neutrinos in most of the right-handed neutrino models. After the Higgs field gets a vacuum expectation value (vev),
\begin{eqnarray}
\left\langle H \right\rangle = \left[
\begin{array}{c}
0 \\
v
\end{array}\right],
\end{eqnarray}
where $v = 174 \text{ GeV}$, the last term in (\ref{NLagrangian}) introduces tiny mixing between $N$ and the SM neutrinos, resulting in effective $N$-$l_i^{\pm}$-$W^{\mp}$, $N$-$\nu_i$-$Z$ vertices through mixing between $N$ and $\nu_i$,
\begin{eqnarray}
\mathcal{L} \supset c{g_2}{\sqrt{2}} \theta_i (W_{\mu}^{+} \bar{N} \gamma^{\mu} P_L l^{-}_i + \text{h.c.}) + \frac{g_2}{\cos \theta_W} \theta_i Z_{\mu} (\bar{N} \gamma^{\mu} P_L \nu_i + \text{h.c.}), \label{N_Decay_Coupling}
\end{eqnarray}
where
\begin{eqnarray}
\theta_i \approx \frac{y_i v}{m_N} \label{MixingParameters}
\end{eqnarray}
are the mixing parameters, $g_2 = \sqrt{ 4 \sqrt{2} G_F m_W^2 }$ is the weak-coupling constant, and $\theta_W$ is the Weinberg angle. Note that (\ref{MixingParameters}) are the first-order results calculated by the perturbation theory of diagonalizing the matrices. As we have noted in the previous section, we only discuss the $m_N < m_W$ case in the following text of this paper, so the effective $N$-$W^{\pm}$-$l^{\mp}$ and $N$-$Z$-$\nu_{e, \mu, \tau}$ coupling constants are similar with the $N$-$H$-$\nu_{e, \mu, \tau}$ coupling constants $y_i$. However, compared with the $W$ and $Z$ bosons, the coupling constants between the SM-Higgs and other light SM fermions are quite small, and all the mediators $Z/W^{\pm}/h$ are off-shell when the right-handed neutrino $N$ decays, so the processes $N \rightarrow h^{*} \nu_i \rightarrow \text{all} + \nu_i$ are negligible. Although the constants $\frac{y_i}{m_N}$ decide the total width of $N$, from (\ref{N_Decay_Coupling}, \ref{MixingParameters}) we can learn that the ratio between the $Z$ and $W^{\pm}$ coupling constants are not affected. That is to say, for each $i=1,2,3$, the ratios $\frac{Br(N \rightarrow \nu_i Z^*)}{Br(N \rightarrow l_{i}^{\pm} W^{\mp *})}$ are fixed if only a $m_N$ is determined.

Assuming that the SM-neutrino masses originate from the Type-I see-saw mechanisms, the mixing parameters $\theta_i \approx \frac{y_i v}{m_N}$ can be large enough so that the lifetime of the right-handed neutrino can be short enough if only $m_N \gtrsim 1 \text{ GeV}$. According to the oscillation data\cite{PDG}, at least one light neutrino should be heavier than $0.1 \text{ eV}$ which means at least one $\frac{y_i^2 v^2}{m_N} > 0.1 \text{ eV}$. If, for example, $m_N = 5 \text{ GeV}$, the mixing parameter $\theta_i = \frac{y_i v}{m_N} > 3 \times 10^{-6}$, which leaves more than enough room beyond the ability of the searching proposals on colliders or other techniques (For some recent experimental and theoretical works on this topic, see Ref.~\cite{RHNCollider1, RHNCollider2, RHNCollider3, RHNCollider4, RHNCollider5, RHNCollider6, RequiredRHNLimit1, RequiredRHNLimit2, RequiredRHNLimit3}). Simple simulations by MadGraph also show that $\tau_{N} \lesssim 10^{-3} \text{ sec} \ll 1 \text{ sec}$, which means once produced, the right-handed neutrinos decay immediately before travelling too far away from the galactic center where they are produced. Further more, we know there are models\cite{Inverse1, Inverse2, Inverse3, Inverse4, ZhouYeLing} that can reach larger $\theta_i$ while keeping light neutrino masses to be small.

\section{Simulations and numerical results}

In this paper, we ignore the inverse Compton, the synchrotron, and the bremsstrahlung emissions from the charged particle products. We only consider the photons emitted during the showering processes and from the decays of the hadrons. Since the decays of the tau leptons produce photons while the electrons and muons do not, we only discuss the following two scenarios for simplicity,
\begin{itemize}
\item $y_1=y_2=0$, $y_3 \neq 0$. Then $100\%$ of the right-handed neutrinos decay through $\tau+W^*$/$\nu_{\tau}+Z^*$ channels. The tau leptons also contribute to the gamma-ray spectrum.
\item $y_3 = 0$, $y_1^2 + y_2^2 \neq 0$. Since muons and electrons do not produce photons, and the ratios $\frac{Br(N \rightarrow \nu_e Z^*)}{Br(N \rightarrow l_{e}^{\pm} W^{\mp *})} = \frac{Br(N \rightarrow \nu_\mu Z^*)}{Br(N \rightarrow l_{\mu}^{\pm} W^{\mp *})}$ are fixed at a given $m_N \gg m_{\mu}$, the gamma-ray spectrum should be independent on concrete values of $y_{1,2}$.
\end{itemize}
The gamma-ray spectrum by general values of $y_{1,2,3}$ are just linear-combinations of the above two cases.

Since we are discussing a pair of dark matter particles $\chi$ annihilating into a pair of right-handed neutrinos $N$, $m_{\chi} \geq m_N$ should be satisfied. If $m_{\chi} > m_N$, the spectrum will also be boosted. The dark matter profile and the $\langle \sigma v \rangle$ also affect the height of the spectrum. In this paper, in order to compare our results with the Ref.~\cite{GCE12, VariousChannel}, we adopt the Navarro-Frenk-White (NFW) profile\cite{NFW},
\begin{eqnarray}
\rho(r) = \rho(r_0) \frac{(r/r_s)^{-\gamma}}{(1+r/r_s)^{3-\gamma}},
\end{eqnarray}
where we adopt $r_s = 20 \text{ kpc}$, $\gamma = 1.2$, and $\rho_0$ is set in order for the local dark matter density $\rho_{\odot}$ to be $0.4 \text{ GeV}/\text{cm}^3$ at $r_{\odot} = 8.5 \text{ kpc}$. The differential flux of the photons from a given direction $\psi$ is given by
\begin{eqnarray}
\frac{dN}{d \Omega dE} (\psi) = \frac{1}{4 \pi \eta} \frac{J(\psi)}{m_{\chi}^2} \langle \sigma v \rangle \frac{dN}{dE_{\gamma}},
\end{eqnarray}
with $\eta = 2(4)$ for the self-conjugate (non-self-conjugate) dark matter. For simplicity and without loss of generality, we adopt $\eta = 2$ in this paper. $\frac{dN}{dE}$ indicates the spectrum of photons emitted per annihilation process. The definition of $J(\psi)$ is given by
\begin{eqnarray}
J(\psi) = \int_{\text{l.o.s}} ds \rho(r)^2
\end{eqnarray}
which is the line-of-sight integral. We use the data integrated within the region of interest (R.O.I) at galactic latitudes $2^{\circ} \leq |b| \leq 20^{\circ}$ and the galactic longitudes $|l| < 20^{\circ}$. The averaged $\bar{J}$ is then to become
\begin{eqnarray}
\bar{J} = \frac{1}{\Delta \Omega} \int_{\Delta \Omega} J(\psi) d\Omega.
\end{eqnarray}
$\bar{J}$ is calculated to be $2.0 \times 10^{23} \text{ GeV}^2/\text{cm}^5$. More realistically, any modifications to the above profile parameters will result in another $\bar{J}_{\text{real}}$. Define
\begin{eqnarray}
\bar{J}_{\text{real}} = \mathcal{J} \times \bar{J},
\end{eqnarray}
then
\begin{eqnarray}
\langle \sigma v \rangle = \frac{\langle \sigma v \rangle_{\text{real}}}{\mathcal{J}},
\end{eqnarray}
where $\langle \sigma v \rangle_{\text{real}}$ is the modified annihilation cross section in this case.

Since there are large correlations among the systematic errors of different bins in the  Calore, Cholis and Weniger's (CCW) fit from the Ref.~\cite{GCE12}, the $\chi^2$ should be defined as
\begin{eqnarray}
\chi^2 = \left[ \frac{dN}{dE} - \left( \frac{dN}{dE} \right)_{\text{obs}} \right] \cdot \Sigma^{-1} \cdot \left[ \frac{dN}{dE} - \left( \frac{dN}{dE} \right)_{\text{obs}} \right].
\end{eqnarray}

\begin{figure}
\includegraphics[width=2.8in]{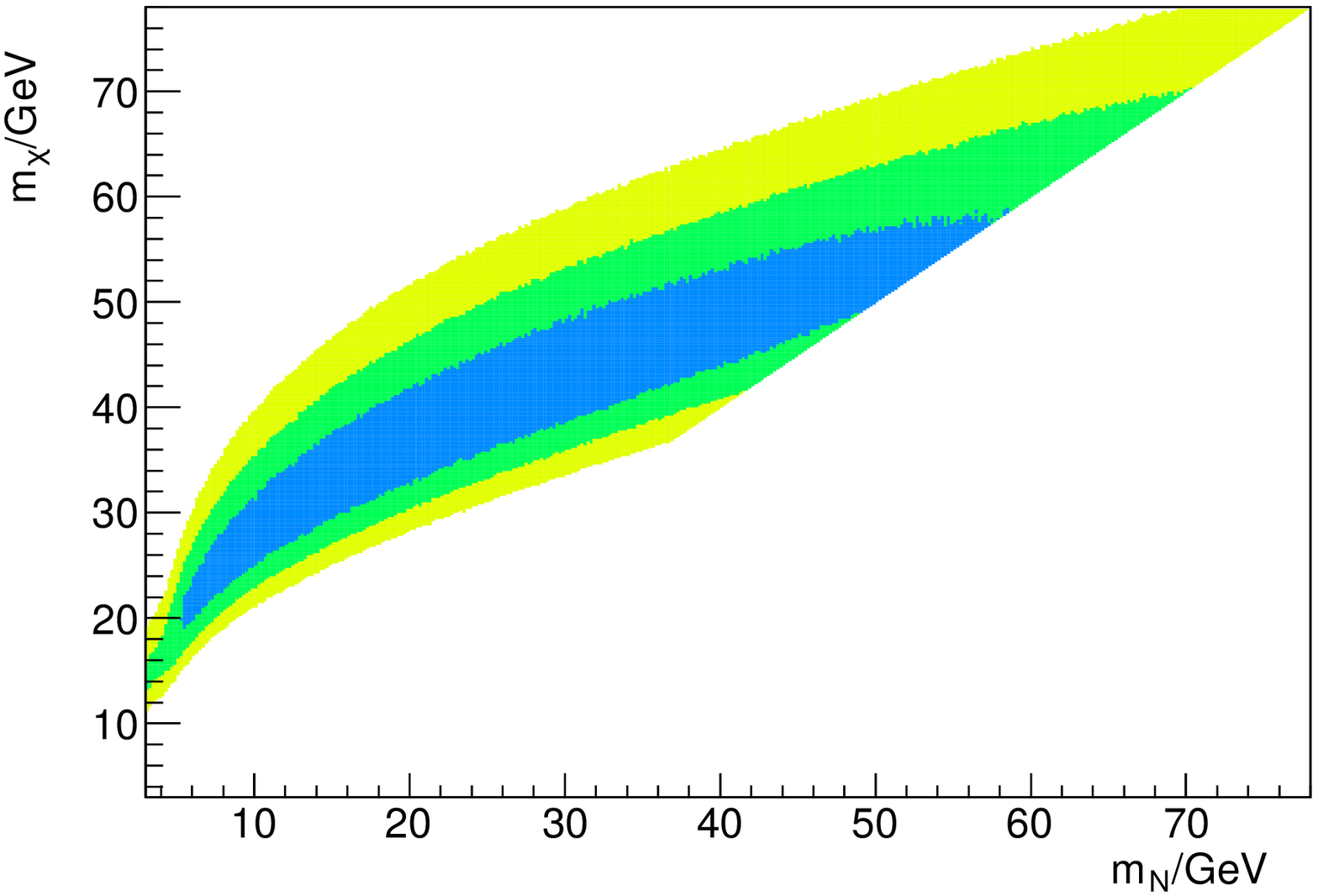}
\includegraphics[width=2.8in]{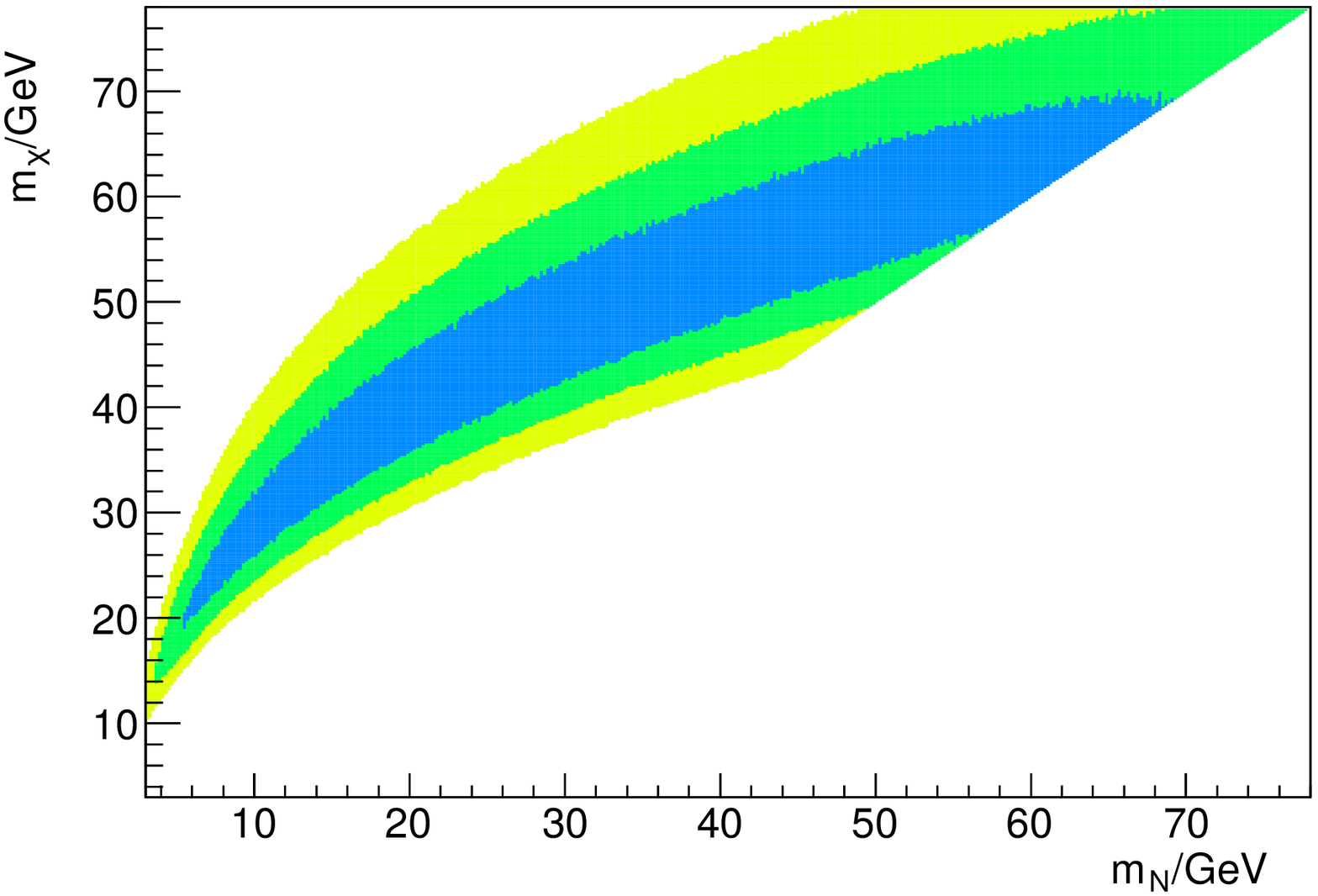}
\caption{The $\Delta \chi^2$ figures. The blue, green, yellow areas are corresponding to the 1,2 and 3 $\sigma$ areas respectively. $\langle \sigma v \rangle$ is adjusted in order to acquire the best-fitted result. The left panel indicates the $y_1=y_2=0$, $y_3 \neq 0$ case. The right-panel indicates the $y_3 = 0$, $y_1^2+y_2^2 \neq 0$ case. \label{Sigma_Contour}}
\end{figure}

\begin{figure}
\includegraphics[width=2.8in]{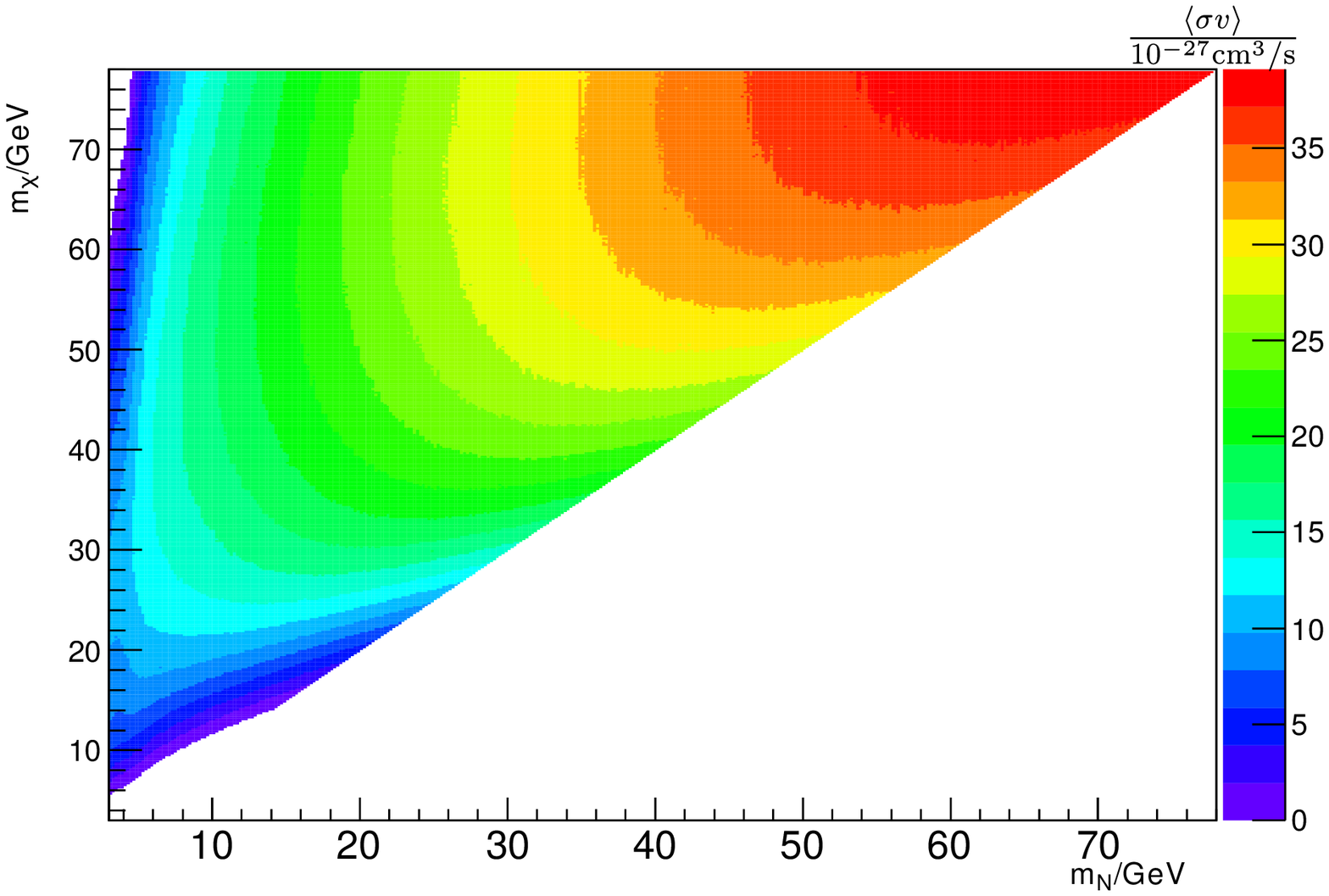}
\includegraphics[width=2.8in]{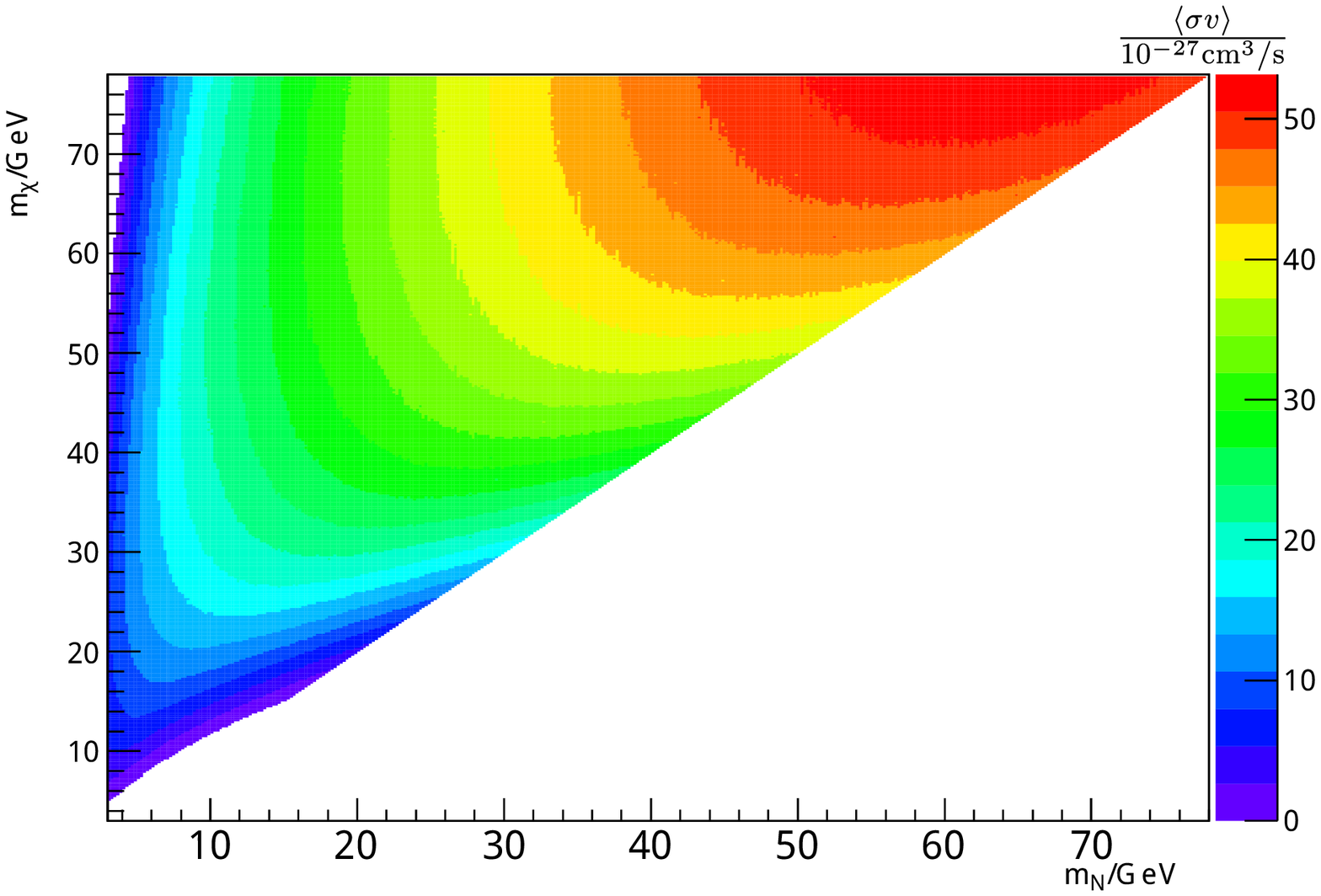}
\caption{The best-fitted $\langle \sigma v \rangle = \langle \sigma v \rangle_{\text{real}} \mathcal{J}$, in the unit of $\text{cm}^3/\text{s}$. The left panel indicates the $y_1=y_2=0$, $y_3 \neq 0$ case. The right-panel indicates the $y_3 = 0$, $y_1^2+y_2^2 \neq 0$ case. \label{Sigmav_Figure}}
\end{figure}

We have scanned the $m_N$-$m_{\chi}$ parameter space by a 0.2 GeV interval. For each point in the parameter space, we used MadGraph5\_aMC@NLO2.3.2 to generate an one-million-event sample file. Then we sent these events to PYTHIA~8.212 in order to acquire the photon spectrum. This process is most time-consuming during the calculations. We list the 1,2 and 3$\sigma$ area in the Fig.~\ref{Sigma_Contour}. The best-fitted points are $m_N = 32.0 \text{ GeV}$, $m_{\chi} = 44.2 \text{ GeV}$, with $\chi^2 = 24.22$ and the best-fitted $\langle \sigma v \rangle = 2.63 \times 10^{-26} \text{cm}^3/\text{s}$ for the $y_1=y_2=0$, $y_3 \neq 0$ case, and $m_N = 27.0 \text{ GeV}$, $m_{\chi} = 45.4 \text{ GeV}$, with $\chi^2 = 23.81$ and the best-fitted $\langle \sigma v \rangle = 3.37 \times 10^{-26} \text{cm}^3/\text{s}$ for the $y_3=0$, $y_1^2+y_2^2 \neq 0$ case. Note that for the $m_N < 10 \text{ GeV}$ cases, which are too near to the $\Lambda_{\text{QCD}}$ scale, the showering and hadronization process from PYTHIA are suspectable. Nevertheless, in the Fig.~\ref{Sigma_Contour}, \ref{Sigmav_Figure}, we still show our numerical results in this area. However, since the ``1-$\sigma$'' area is such a long belt ranging from 10 GeV to 60 GeV, the main features of our conclusions should not be affected severely by the uncertainty of the QCD calculations.

In the Fig.~\ref{Sigmav_Figure}, we also plot the best-fitted $\langle \sigma v \rangle$ for each $m_{N}$ and $m_{\chi}$.

\begin{figure}
\includegraphics[width=4.5in]{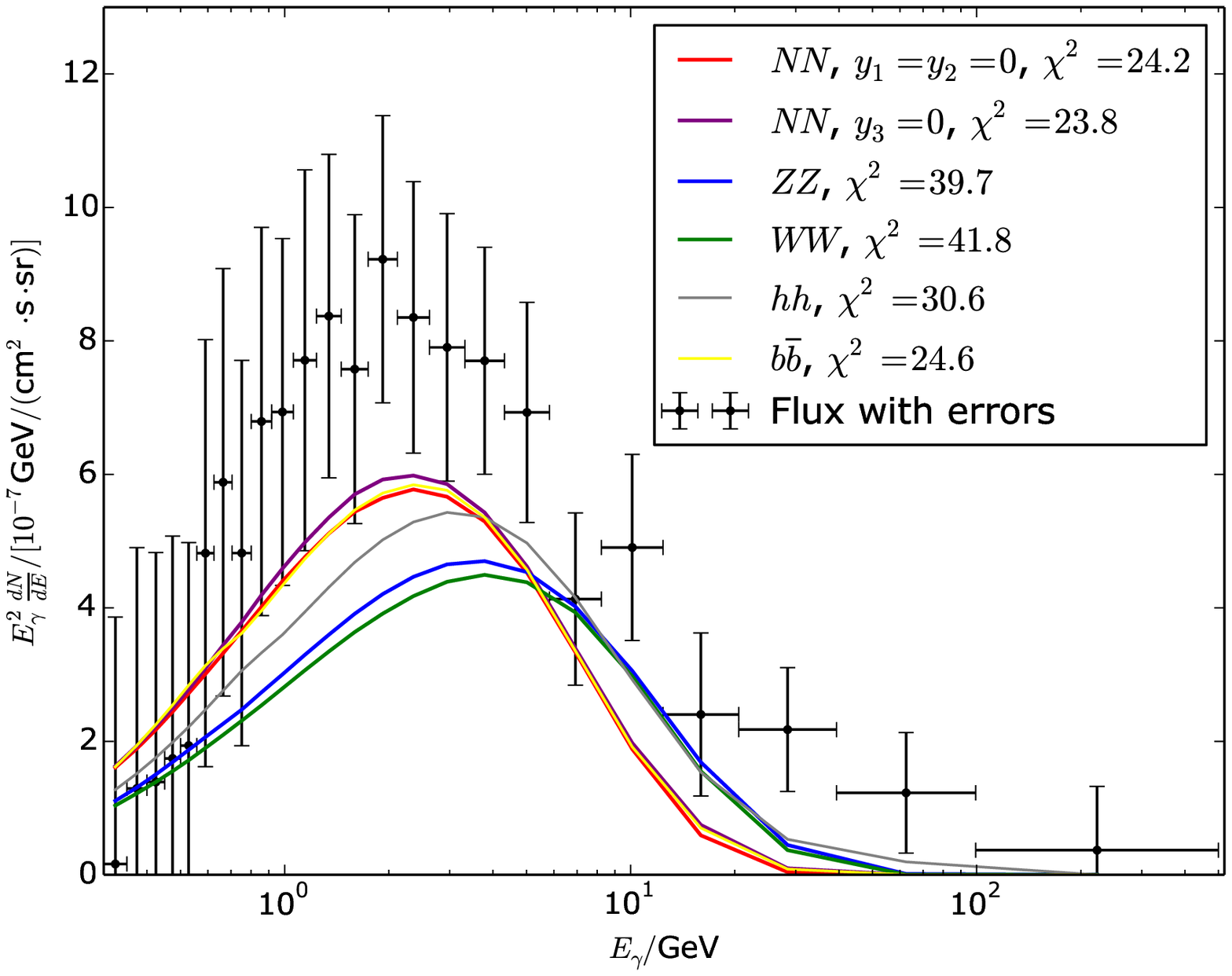}
\caption{The best-fitted gamma-ray spectrum together with the observed central values and the errorbars. In the case of $y_1=y_2=0$, $y_3 \neq 0$, $\chi^2 = 24.22$, with the p-value $0.336$. In the case of $y_3 = 0$, $y_1^2+y_2^2 \neq 0$, $\chi^2 = 23.81$, with p-value $0.357$. The data together with the error bars are from Ref.~\cite{GCE12}. We also plot the gamma-ray spectrum and list the $\chi^2$ value of the best-fitted $ZZ$, $WW$, $hh$, and $b \overline{b}$ channels for comparison. All these curves and values are calculated by a similar MadGraph5\_aMC@NLO2.3.2+PYTHIA~8.212 process.
\label{Spectrum}}
\end{figure}

From the Fig.~\ref{Sigma_Contour}, \ref{Sigmav_Figure}, and \ref{Spectrum} we can learn that $m_N$ approximately ranges from $10 \text{ GeV}$ to $60 \text{ GeV}$ within 1 $\sigma$ level and necessary boost is needed for the best-fitting with the observed excess. Both the $y_1=y_2=0$, $y_3 \neq 0$, and $y_3 = 0$, $y_1^2+y_2^2 \neq 0$ show us no significant difference between them. However, a slightly larger $\langle \sigma v \rangle$ is needed in the $y_3 = 0$, $y_1^2 + y_2^2 \neq 0$ case. This is because the tau leptons produced in the $y_3 \neq 0$ case strengthen the gamma-ray flux, thus weaken the needed $\langle \sigma v \rangle$ in this case.

Since the gamma-ray signals from near the center of our galaxy are severely contaminated, it is quite important to compare the results with the constraints from the dSph galaxies. Detailed analyses depend closely on the shapes and the fluxes of the spectrum in specific models (For an example of the method, see Ref.~\cite{Fittingdsph}). In this paper, we only note that from observing the spectrum depicted in Fig.~\ref{Spectrum}, the best-fitted spectrum induced from the right-handed neutrinos are fairly close to the ones induced from the $b \overline{b}$ channel which can help us roughly infer the situation of the right-handed neutrino. According to the Ref.~\cite{dSph7}, the best-fitted $b \overline{b}$ point is located just slightly above the constraint line, which is a subtle case. However, the annihilation rate $\langle \sigma v \rangle_{\text{real}} = \mathcal{J} \langle \sigma v \rangle$ is quite sensitive to the profile parameters, where relatively large uncertainties remain \cite{ProfileUncertainty}. For example, if the local dark matter density $\rho_{\odot}$ varies from $0.2 \text{ GeV}/\text{cm}^3$ to $0.6 \text{ GeV}/\text{cm}^3$, $\langle \sigma v \rangle_{\text{real}}$ can differ by one order of magnitude since it depends on $\rho_{\odot}^2$. In this sense, the results from near our galactic center and the constraints from the dSph galaxies are still compatible at present stage.

\section{Summary and Conclusions}

In this paper, we calculated the predicted gamma-ray excess from the galactic center in a specific case that dark matter particles annihilate into two light right-handed neutrinos. We find that the $m_N$ can range from approximately $10 \text{ GeV}$ to $60 \text{ GeV}$ within 1 $\sigma$ level and necessary boost is needed. The $\langle \sigma v \rangle$ can vary from $0.5 \times 10^{-26} \text{cm}^3/\text{s}$ to $5 \times 10^{-26} \text{cm}^3/\text{s}$, which is roughly compatible with the WIMP annihilation cross section $\langle \sigma v \rangle_{\text{decouple}} = 2\text{-}3 \times 10^{-26}\text{cm}^3/\text{s}$ when the dark matter particles decouple. Comparing the two panels in the Fig.~\ref{Sigmav_Figure}, we can see that we need a slightly larger $\langle \sigma v \rangle$ in the $y_3 = 0$, $y_1^2 + y_2^2 \neq 0$ case, since the tau leptons produced in this case is much less then the $y_3 \neq 0$ case.

In this paper, we extract some common points from some specific new physics models in which the DM might annihilate mainly into right-handed neutrinos. The detailed properties of the dark matter particles are not discussed in this paper, and the right-handed neutrinos are so weakly coupled with the SM sectors, so it is hardly possible for us to test this scenario by other ways, e.g., collider physics. However, in these specific models in which this scenario is embedded, such right-handed neutrinos can be produced through other mediators. For example, in the NMSSM extended with the right-handed neutrino(s), the exotic singlet-like Higgs boson might decay into the right-handed neutrinos. Detailed discussions about these models in this case shall also be our next topics.

\begin{acknowledgements}

We would like to thank Ran Ding, Jia-Shu Lu, Weicong Huang, Chen Zhang, Weihong Zhang, Yu-Feng Zhou, Lei Feng for helpful discussions.  This work was supported in part by the Natural Science Foundation of China (Grants No.~11135003 and No.~11375014).

\end{acknowledgements}

\newpage
\bibliography{DMRHN}

\begin{thebibliography}{82}
\expandafter\ifx\csname natexlab\endcsname\relax\def\natexlab#1{#1}\fi
\expandafter\ifx\csname bibnamefont\endcsname\relax
  \def\bibnamefont#1{#1}\fi
\expandafter\ifx\csname bibfnamefont\endcsname\relax
  \def\bibfnamefont#1{#1}\fi
\expandafter\ifx\csname citenamefont\endcsname\relax
  \def\citenamefont#1{#1}\fi
\expandafter\ifx\csname url\endcsname\relax
  \def\url#1{\texttt{#1}}\fi
\expandafter\ifx\csname urlprefix\endcsname\relax\def\urlprefix{URL }\fi
\providecommand{\bibinfo}[2]{#2}
\providecommand{\eprint}[2][]{\url{#2}}

\bibitem[{\citenamefont{Goodenough and Hooper}(2009)}]{GCE1}
\bibinfo{author}{\bibfnamefont{L.}~\bibnamefont{Goodenough}} \bibnamefont{and}
  \bibinfo{author}{\bibfnamefont{D.}~\bibnamefont{Hooper}}
  (\bibinfo{year}{2009}), \eprint{0910.2998}.

\bibitem[{\citenamefont{Hooper and Goodenough}(2011)}]{GCE2}
\bibinfo{author}{\bibfnamefont{D.}~\bibnamefont{Hooper}} \bibnamefont{and}
  \bibinfo{author}{\bibfnamefont{L.}~\bibnamefont{Goodenough}},
  \bibinfo{journal}{Phys. Lett.} \textbf{\bibinfo{volume}{B697}},
  \bibinfo{pages}{412} (\bibinfo{year}{2011}), \eprint{1010.2752}.

\bibitem[{\citenamefont{Boyarsky et~al.}(2011)\citenamefont{Boyarsky, Malyshev,
  and Ruchayskiy}}]{GCE3}
\bibinfo{author}{\bibfnamefont{A.}~\bibnamefont{Boyarsky}},
  \bibinfo{author}{\bibfnamefont{D.}~\bibnamefont{Malyshev}}, \bibnamefont{and}
  \bibinfo{author}{\bibfnamefont{O.}~\bibnamefont{Ruchayskiy}},
  \bibinfo{journal}{Phys. Lett.} \textbf{\bibinfo{volume}{B705}},
  \bibinfo{pages}{165} (\bibinfo{year}{2011}), \eprint{1012.5839}.

\bibitem[{\citenamefont{Hooper and Linden}(2011)}]{GCE4}
\bibinfo{author}{\bibfnamefont{D.}~\bibnamefont{Hooper}} \bibnamefont{and}
  \bibinfo{author}{\bibfnamefont{T.}~\bibnamefont{Linden}},
  \bibinfo{journal}{Phys. Rev.} \textbf{\bibinfo{volume}{D84}},
  \bibinfo{pages}{123005} (\bibinfo{year}{2011}), \eprint{1110.0006}.

\bibitem[{\citenamefont{Linden et~al.}(2012)\citenamefont{Linden, Lovegrove,
  and Profumo}}]{GCE5}
\bibinfo{author}{\bibfnamefont{T.}~\bibnamefont{Linden}},
  \bibinfo{author}{\bibfnamefont{E.}~\bibnamefont{Lovegrove}},
  \bibnamefont{and} \bibinfo{author}{\bibfnamefont{S.}~\bibnamefont{Profumo}},
  \bibinfo{journal}{Astrophys. J.} \textbf{\bibinfo{volume}{753}},
  \bibinfo{pages}{41} (\bibinfo{year}{2012}), \eprint{1203.3539}.

\bibitem[{\citenamefont{Abazajian and Kaplinghat}(2012)}]{GCE6}
\bibinfo{author}{\bibfnamefont{K.~N.} \bibnamefont{Abazajian}}
  \bibnamefont{and}
  \bibinfo{author}{\bibfnamefont{M.}~\bibnamefont{Kaplinghat}},
  \bibinfo{journal}{Phys. Rev.} \textbf{\bibinfo{volume}{D86}},
  \bibinfo{pages}{083511} (\bibinfo{year}{2012}), \bibinfo{note}{[Erratum:
  Phys. Rev.D87,129902(2013)]}, \eprint{1207.6047}.

\bibitem[{\citenamefont{Hooper and Slatyer}(2013)}]{GCE7}
\bibinfo{author}{\bibfnamefont{D.}~\bibnamefont{Hooper}} \bibnamefont{and}
  \bibinfo{author}{\bibfnamefont{T.~R.} \bibnamefont{Slatyer}},
  \bibinfo{journal}{Phys. Dark Univ.} \textbf{\bibinfo{volume}{2}},
  \bibinfo{pages}{118} (\bibinfo{year}{2013}), \eprint{1302.6589}.

\bibitem[{\citenamefont{Gordon and Macias}(2013)}]{GCE8}
\bibinfo{author}{\bibfnamefont{C.}~\bibnamefont{Gordon}} \bibnamefont{and}
  \bibinfo{author}{\bibfnamefont{O.}~\bibnamefont{Macias}},
  \bibinfo{journal}{Phys. Rev.} \textbf{\bibinfo{volume}{D88}},
  \bibinfo{pages}{083521} (\bibinfo{year}{2013}), \bibinfo{note}{[Erratum:
  Phys. Rev.D89,no.4,049901(2014)]}, \eprint{1306.5725}.

\bibitem[{\citenamefont{Abazajian et~al.}(2014)\citenamefont{Abazajian, Canac,
  Horiuchi, and Kaplinghat}}]{GCE9}
\bibinfo{author}{\bibfnamefont{K.~N.} \bibnamefont{Abazajian}},
  \bibinfo{author}{\bibfnamefont{N.}~\bibnamefont{Canac}},
  \bibinfo{author}{\bibfnamefont{S.}~\bibnamefont{Horiuchi}}, \bibnamefont{and}
  \bibinfo{author}{\bibfnamefont{M.}~\bibnamefont{Kaplinghat}},
  \bibinfo{journal}{Phys. Rev.} \textbf{\bibinfo{volume}{D90}},
  \bibinfo{pages}{023526} (\bibinfo{year}{2014}), \eprint{1402.4090}.

\bibitem[{\citenamefont{Daylan et~al.}(2014)\citenamefont{Daylan, Finkbeiner,
  Hooper, Linden, Portillo, Rodd, and Slatyer}}]{GCE10}
\bibinfo{author}{\bibfnamefont{T.}~\bibnamefont{Daylan}},
  \bibinfo{author}{\bibfnamefont{D.~P.} \bibnamefont{Finkbeiner}},
  \bibinfo{author}{\bibfnamefont{D.}~\bibnamefont{Hooper}},
  \bibinfo{author}{\bibfnamefont{T.}~\bibnamefont{Linden}},
  \bibinfo{author}{\bibfnamefont{S.~K.~N.} \bibnamefont{Portillo}},
  \bibinfo{author}{\bibfnamefont{N.~L.} \bibnamefont{Rodd}}, \bibnamefont{and}
  \bibinfo{author}{\bibfnamefont{T.~R.} \bibnamefont{Slatyer}}
  (\bibinfo{year}{2014}), \eprint{1402.6703}.

\bibitem[{\citenamefont{Zhou et~al.}(2015)\citenamefont{Zhou, Liang, Huang, Li,
  Fan, Feng, and Chang}}]{GCE11}
\bibinfo{author}{\bibfnamefont{B.}~\bibnamefont{Zhou}},
  \bibinfo{author}{\bibfnamefont{Y.-F.} \bibnamefont{Liang}},
  \bibinfo{author}{\bibfnamefont{X.}~\bibnamefont{Huang}},
  \bibinfo{author}{\bibfnamefont{X.}~\bibnamefont{Li}},
  \bibinfo{author}{\bibfnamefont{Y.-Z.} \bibnamefont{Fan}},
  \bibinfo{author}{\bibfnamefont{L.}~\bibnamefont{Feng}}, \bibnamefont{and}
  \bibinfo{author}{\bibfnamefont{J.}~\bibnamefont{Chang}},
  \bibinfo{journal}{Phys. Rev.} \textbf{\bibinfo{volume}{D91}},
  \bibinfo{pages}{123010} (\bibinfo{year}{2015}), \eprint{1406.6948}.

\bibitem[{\citenamefont{Calore et~al.}(2015)\citenamefont{Calore, Cholis, and
  Weniger}}]{GCE12}
\bibinfo{author}{\bibfnamefont{F.}~\bibnamefont{Calore}},
  \bibinfo{author}{\bibfnamefont{I.}~\bibnamefont{Cholis}}, \bibnamefont{and}
  \bibinfo{author}{\bibfnamefont{C.}~\bibnamefont{Weniger}},
  \bibinfo{journal}{JCAP} \textbf{\bibinfo{volume}{1503}}, \bibinfo{pages}{038}
  (\bibinfo{year}{2015}), \eprint{1409.0042}.

\bibitem[{\citenamefont{Agrawal et~al.}(2015)\citenamefont{Agrawal, Batell,
  Fox, and Harnik}}]{VariousChannel}
\bibinfo{author}{\bibfnamefont{P.}~\bibnamefont{Agrawal}},
  \bibinfo{author}{\bibfnamefont{B.}~\bibnamefont{Batell}},
  \bibinfo{author}{\bibfnamefont{P.~J.} \bibnamefont{Fox}}, \bibnamefont{and}
  \bibinfo{author}{\bibfnamefont{R.}~\bibnamefont{Harnik}},
  \bibinfo{journal}{JCAP} \textbf{\bibinfo{volume}{1505}}, \bibinfo{pages}{011}
  (\bibinfo{year}{2015}), \eprint{1411.2592}.

\bibitem[{\citenamefont{Alves et~al.}(2014)\citenamefont{Alves, Profumo,
  Queiroz, and Shepherd}}]{WhereToPut1}
\bibinfo{author}{\bibfnamefont{A.}~\bibnamefont{Alves}},
  \bibinfo{author}{\bibfnamefont{S.}~\bibnamefont{Profumo}},
  \bibinfo{author}{\bibfnamefont{F.~S.} \bibnamefont{Queiroz}},
  \bibnamefont{and} \bibinfo{author}{\bibfnamefont{W.}~\bibnamefont{Shepherd}},
  \bibinfo{journal}{Phys. Rev.} \textbf{\bibinfo{volume}{D90}},
  \bibinfo{pages}{115003} (\bibinfo{year}{2014}), \eprint{1403.5027}.

\bibitem[{\citenamefont{Duerr et~al.}(2015)\citenamefont{Duerr, Fileviez~Perez,
  and Smirnov}}]{WhereToPut2}
\bibinfo{author}{\bibfnamefont{M.}~\bibnamefont{Duerr}},
  \bibinfo{author}{\bibfnamefont{P.}~\bibnamefont{Fileviez~Perez}},
  \bibnamefont{and} \bibinfo{author}{\bibfnamefont{J.}~\bibnamefont{Smirnov}}
  (\bibinfo{year}{2015}), \eprint{1510.07562}.

\bibitem[{\citenamefont{Ruiz-Alvarez et~al.}(2012)\citenamefont{Ruiz-Alvarez,
  de~S.~Pires, Queiroz, Restrepo, and Rodrigues~da Silva}}]{WhereToPut3}
\bibinfo{author}{\bibfnamefont{J.~D.} \bibnamefont{Ruiz-Alvarez}},
  \bibinfo{author}{\bibfnamefont{C.~A.} \bibnamefont{de~S.~Pires}},
  \bibinfo{author}{\bibfnamefont{F.~S.} \bibnamefont{Queiroz}},
  \bibinfo{author}{\bibfnamefont{D.}~\bibnamefont{Restrepo}}, \bibnamefont{and}
  \bibinfo{author}{\bibfnamefont{P.~S.} \bibnamefont{Rodrigues~da Silva}},
  \bibinfo{journal}{Phys. Rev.} \textbf{\bibinfo{volume}{D86}},
  \bibinfo{pages}{075011} (\bibinfo{year}{2012}), \eprint{1206.5779}.

\bibitem[{\citenamefont{Cao et~al.}(2015{\natexlab{a}})\citenamefont{Cao,
  Shang, Wu, Yang, and Zhang}}]{WhereToPut4}
\bibinfo{author}{\bibfnamefont{J.}~\bibnamefont{Cao}},
  \bibinfo{author}{\bibfnamefont{L.}~\bibnamefont{Shang}},
  \bibinfo{author}{\bibfnamefont{P.}~\bibnamefont{Wu}},
  \bibinfo{author}{\bibfnamefont{J.~M.} \bibnamefont{Yang}}, \bibnamefont{and}
  \bibinfo{author}{\bibfnamefont{Y.}~\bibnamefont{Zhang}},
  \bibinfo{journal}{Phys. Rev.} \textbf{\bibinfo{volume}{D91}},
  \bibinfo{pages}{055005} (\bibinfo{year}{2015}{\natexlab{a}}),
  \eprint{1410.3239}.

\bibitem[{\citenamefont{Ghorbani}(2015)}]{WhereToPut5}
\bibinfo{author}{\bibfnamefont{K.}~\bibnamefont{Ghorbani}},
  \bibinfo{journal}{JCAP} \textbf{\bibinfo{volume}{1501}}, \bibinfo{pages}{015}
  (\bibinfo{year}{2015}), \eprint{1408.4929}.

\bibitem[{\citenamefont{Ghorbani and Ghorbani}(2014)}]{WhereToPut6}
\bibinfo{author}{\bibfnamefont{K.}~\bibnamefont{Ghorbani}} \bibnamefont{and}
  \bibinfo{author}{\bibfnamefont{H.}~\bibnamefont{Ghorbani}}
  (\bibinfo{year}{2014}), \eprint{1501.00206}.

\bibitem[{\citenamefont{Ghorbani and Ghorbani}(2015)}]{WhereToPut7}
\bibinfo{author}{\bibfnamefont{K.}~\bibnamefont{Ghorbani}} \bibnamefont{and}
  \bibinfo{author}{\bibfnamefont{H.}~\bibnamefont{Ghorbani}},
  \bibinfo{journal}{Phys. Rev.} \textbf{\bibinfo{volume}{D91}},
  \bibinfo{pages}{123541} (\bibinfo{year}{2015}), \eprint{1504.03610}.

\bibitem[{\citenamefont{Ackermann et~al.}(2011)}]{dSph1}
\bibinfo{author}{\bibfnamefont{M.}~\bibnamefont{Ackermann}}
  \bibnamefont{et~al.} (\bibinfo{collaboration}{Fermi-LAT}),
  \bibinfo{journal}{Phys. Rev. Lett.} \textbf{\bibinfo{volume}{107}},
  \bibinfo{pages}{241302} (\bibinfo{year}{2011}), \eprint{1108.3546}.

\bibitem[{\citenamefont{Geringer-Sameth and Koushiappas}(2011)}]{dSph2}
\bibinfo{author}{\bibfnamefont{A.}~\bibnamefont{Geringer-Sameth}}
  \bibnamefont{and} \bibinfo{author}{\bibfnamefont{S.~M.}
  \bibnamefont{Koushiappas}}, \bibinfo{journal}{Phys. Rev. Lett.}
  \textbf{\bibinfo{volume}{107}}, \bibinfo{pages}{241303}
  (\bibinfo{year}{2011}), \eprint{1108.2914}.

\bibitem[{\citenamefont{Tsai et~al.}(2013{\natexlab{a}})\citenamefont{Tsai,
  Yuan, and Huang}}]{dSph3}
\bibinfo{author}{\bibfnamefont{Y.-L.~S.} \bibnamefont{Tsai}},
  \bibinfo{author}{\bibfnamefont{Q.}~\bibnamefont{Yuan}}, \bibnamefont{and}
  \bibinfo{author}{\bibfnamefont{X.}~\bibnamefont{Huang}},
  \bibinfo{journal}{JCAP} \textbf{\bibinfo{volume}{1303}}, \bibinfo{pages}{018}
  (\bibinfo{year}{2013}{\natexlab{a}}), \eprint{1212.3990}.

\bibitem[{\citenamefont{Mazziotta et~al.}(2012)\citenamefont{Mazziotta,
  Loparco, de~Palma, and Giglietto}}]{dSph4}
\bibinfo{author}{\bibfnamefont{M.~N.} \bibnamefont{Mazziotta}},
  \bibinfo{author}{\bibfnamefont{F.}~\bibnamefont{Loparco}},
  \bibinfo{author}{\bibfnamefont{F.}~\bibnamefont{de~Palma}}, \bibnamefont{and}
  \bibinfo{author}{\bibfnamefont{N.}~\bibnamefont{Giglietto}},
  \bibinfo{journal}{Astropart. Phys.} \textbf{\bibinfo{volume}{37}},
  \bibinfo{pages}{26} (\bibinfo{year}{2012}), \eprint{1203.6731}.

\bibitem[{\citenamefont{Ackermann et~al.}(2014)}]{dSph5}
\bibinfo{author}{\bibfnamefont{M.}~\bibnamefont{Ackermann}}
  \bibnamefont{et~al.} (\bibinfo{collaboration}{Fermi-LAT}),
  \bibinfo{journal}{Phys. Rev.} \textbf{\bibinfo{volume}{D89}},
  \bibinfo{pages}{042001} (\bibinfo{year}{2014}), \eprint{1310.0828}.

\bibitem[{\citenamefont{Ackermann et~al.}(2015)}]{dSph6}
\bibinfo{author}{\bibfnamefont{M.}~\bibnamefont{Ackermann}}
  \bibnamefont{et~al.} (\bibinfo{collaboration}{Fermi-LAT}),
  \bibinfo{journal}{Phys. Rev. Lett.} \textbf{\bibinfo{volume}{115}},
  \bibinfo{pages}{231301} (\bibinfo{year}{2015}), \eprint{1503.02641}.

\bibitem[{\citenamefont{Geringer-Sameth
  et~al.}(2015{\natexlab{a}})\citenamefont{Geringer-Sameth, Koushiappas, and
  Walker}}]{dSph7}
\bibinfo{author}{\bibfnamefont{A.}~\bibnamefont{Geringer-Sameth}},
  \bibinfo{author}{\bibfnamefont{S.~M.} \bibnamefont{Koushiappas}},
  \bibnamefont{and} \bibinfo{author}{\bibfnamefont{M.~G.}
  \bibnamefont{Walker}}, \bibinfo{journal}{Phys. Rev.}
  \textbf{\bibinfo{volume}{D91}}, \bibinfo{pages}{083535}
  (\bibinfo{year}{2015}{\natexlab{a}}), \eprint{1410.2242}.

\bibitem[{\citenamefont{Wood et~al.}(2015)\citenamefont{Wood, Anderson,
  Drlica-Wagner, Cohen-Tanugi, and Conrad}}]{dsph8}
\bibinfo{author}{\bibfnamefont{M.}~\bibnamefont{Wood}},
  \bibinfo{author}{\bibfnamefont{B.}~\bibnamefont{Anderson}},
  \bibinfo{author}{\bibfnamefont{A.}~\bibnamefont{Drlica-Wagner}},
  \bibinfo{author}{\bibfnamefont{J.}~\bibnamefont{Cohen-Tanugi}},
  \bibnamefont{and} \bibinfo{author}{\bibfnamefont{J.}~\bibnamefont{Conrad}}
  (\bibinfo{collaboration}{Fermi-LAT}), in
  \emph{\bibinfo{booktitle}{{Proceedings, 34th International Cosmic Ray
  Conference (ICRC 2015)}}} (\bibinfo{year}{2015}), \eprint{1507.03530},
  \urlprefix\url{http://inspirehep.net/record/1382564/files/arXiv:1507.03530.pdf}.

\bibitem[{\citenamefont{Geringer-Sameth
  et~al.}(2015{\natexlab{b}})\citenamefont{Geringer-Sameth, Walker,
  Koushiappas, Koposov, Belokurov, Torrealba, and Evans}}]{dSphDiscover1}
\bibinfo{author}{\bibfnamefont{A.}~\bibnamefont{Geringer-Sameth}},
  \bibinfo{author}{\bibfnamefont{M.~G.} \bibnamefont{Walker}},
  \bibinfo{author}{\bibfnamefont{S.~M.} \bibnamefont{Koushiappas}},
  \bibinfo{author}{\bibfnamefont{S.~E.} \bibnamefont{Koposov}},
  \bibinfo{author}{\bibfnamefont{V.}~\bibnamefont{Belokurov}},
  \bibinfo{author}{\bibfnamefont{G.}~\bibnamefont{Torrealba}},
  \bibnamefont{and} \bibinfo{author}{\bibfnamefont{N.~W.} \bibnamefont{Evans}},
  \bibinfo{journal}{Phys. Rev. Lett.} \textbf{\bibinfo{volume}{115}},
  \bibinfo{pages}{081101} (\bibinfo{year}{2015}{\natexlab{b}}),
  \eprint{1503.02320}.

\bibitem[{\citenamefont{Hooper and Linden}(2015)}]{dSphDiscover2}
\bibinfo{author}{\bibfnamefont{D.}~\bibnamefont{Hooper}} \bibnamefont{and}
  \bibinfo{author}{\bibfnamefont{T.}~\bibnamefont{Linden}},
  \bibinfo{journal}{JCAP} \textbf{\bibinfo{volume}{1509}}, \bibinfo{pages}{016}
  (\bibinfo{year}{2015}), \eprint{1503.06209}.

\bibitem[{\citenamefont{Drlica-Wagner et~al.}(2015)}]{dSphDiscover3}
\bibinfo{author}{\bibfnamefont{A.}~\bibnamefont{Drlica-Wagner}}
  \bibnamefont{et~al.} (\bibinfo{collaboration}{DES, Fermi-LAT}),
  \bibinfo{journal}{Astrophys. J.} \textbf{\bibinfo{volume}{809}},
  \bibinfo{pages}{L4} (\bibinfo{year}{2015}), \eprint{1503.02632}.

\bibitem[{\citenamefont{Li et~al.}(2015)\citenamefont{Li, Liang, Duan, Shen,
  Huang, Li, Fan, Liao, Feng, and Chang}}]{dSphDiscover4}
\bibinfo{author}{\bibfnamefont{S.}~\bibnamefont{Li}},
  \bibinfo{author}{\bibfnamefont{Y.-F.} \bibnamefont{Liang}},
  \bibinfo{author}{\bibfnamefont{K.-K.} \bibnamefont{Duan}},
  \bibinfo{author}{\bibfnamefont{Z.-Q.} \bibnamefont{Shen}},
  \bibinfo{author}{\bibfnamefont{X.}~\bibnamefont{Huang}},
  \bibinfo{author}{\bibfnamefont{X.}~\bibnamefont{Li}},
  \bibinfo{author}{\bibfnamefont{Y.-Z.} \bibnamefont{Fan}},
  \bibinfo{author}{\bibfnamefont{N.-H.} \bibnamefont{Liao}},
  \bibinfo{author}{\bibfnamefont{L.}~\bibnamefont{Feng}}, \bibnamefont{and}
  \bibinfo{author}{\bibfnamefont{J.}~\bibnamefont{Chang}}
  (\bibinfo{year}{2015}), \eprint{1511.09252}.

\bibitem[{\citenamefont{Martin et~al.}(2014)\citenamefont{Martin, Shelton, and
  Unwin}}]{Cascade1}
\bibinfo{author}{\bibfnamefont{A.}~\bibnamefont{Martin}},
  \bibinfo{author}{\bibfnamefont{J.}~\bibnamefont{Shelton}}, \bibnamefont{and}
  \bibinfo{author}{\bibfnamefont{J.}~\bibnamefont{Unwin}},
  \bibinfo{journal}{Phys. Rev.} \textbf{\bibinfo{volume}{D90}},
  \bibinfo{pages}{103513} (\bibinfo{year}{2014}), \eprint{1405.0272}.

\bibitem[{\citenamefont{Cerdeno et~al.}(2015)\citenamefont{Cerdeno, Peiro, and
  Robles}}]{Cascade2NMSSMN}
\bibinfo{author}{\bibfnamefont{D.~G.} \bibnamefont{Cerdeno}},
  \bibinfo{author}{\bibfnamefont{M.}~\bibnamefont{Peiro}}, \bibnamefont{and}
  \bibinfo{author}{\bibfnamefont{S.}~\bibnamefont{Robles}},
  \bibinfo{journal}{Phys. Rev.} \textbf{\bibinfo{volume}{D91}},
  \bibinfo{pages}{123530} (\bibinfo{year}{2015}), \eprint{1501.01296}.

\bibitem[{\citenamefont{Rajaraman et~al.}(2015)\citenamefont{Rajaraman,
  Smolinsky, and Tanedo}}]{Cascade3}
\bibinfo{author}{\bibfnamefont{A.}~\bibnamefont{Rajaraman}},
  \bibinfo{author}{\bibfnamefont{J.}~\bibnamefont{Smolinsky}},
  \bibnamefont{and} \bibinfo{author}{\bibfnamefont{P.}~\bibnamefont{Tanedo}}
  (\bibinfo{year}{2015}), \eprint{1503.05919}.

\bibitem[{\citenamefont{Abdullah et~al.}(2014)\citenamefont{Abdullah, DiFranzo,
  Rajaraman, Tait, Tanedo, and Wijangco}}]{Cascade4}
\bibinfo{author}{\bibfnamefont{M.}~\bibnamefont{Abdullah}},
  \bibinfo{author}{\bibfnamefont{A.}~\bibnamefont{DiFranzo}},
  \bibinfo{author}{\bibfnamefont{A.}~\bibnamefont{Rajaraman}},
  \bibinfo{author}{\bibfnamefont{T.~M.~P.} \bibnamefont{Tait}},
  \bibinfo{author}{\bibfnamefont{P.}~\bibnamefont{Tanedo}}, \bibnamefont{and}
  \bibinfo{author}{\bibfnamefont{A.~M.} \bibnamefont{Wijangco}},
  \bibinfo{journal}{Phys. Rev.} \textbf{\bibinfo{volume}{D90}},
  \bibinfo{pages}{035004} (\bibinfo{year}{2014}), \eprint{1404.6528}.

\bibitem[{\citenamefont{Cao et~al.}(2015{\natexlab{b}})\citenamefont{Cao,
  Shang, Wu, Yang, and Zhang}}]{Cascade5}
\bibinfo{author}{\bibfnamefont{J.}~\bibnamefont{Cao}},
  \bibinfo{author}{\bibfnamefont{L.}~\bibnamefont{Shang}},
  \bibinfo{author}{\bibfnamefont{P.}~\bibnamefont{Wu}},
  \bibinfo{author}{\bibfnamefont{J.~M.} \bibnamefont{Yang}}, \bibnamefont{and}
  \bibinfo{author}{\bibfnamefont{Y.}~\bibnamefont{Zhang}},
  \bibinfo{journal}{JHEP} \textbf{\bibinfo{volume}{10}}, \bibinfo{pages}{030}
  (\bibinfo{year}{2015}{\natexlab{b}}), \eprint{1506.06471}.

\bibitem[{\citenamefont{Berlin et~al.}(2014)\citenamefont{Berlin, Gratia,
  Hooper, and McDermott}}]{Cascade6}
\bibinfo{author}{\bibfnamefont{A.}~\bibnamefont{Berlin}},
  \bibinfo{author}{\bibfnamefont{P.}~\bibnamefont{Gratia}},
  \bibinfo{author}{\bibfnamefont{D.}~\bibnamefont{Hooper}}, \bibnamefont{and}
  \bibinfo{author}{\bibfnamefont{S.~D.} \bibnamefont{McDermott}},
  \bibinfo{journal}{Phys. Rev.} \textbf{\bibinfo{volume}{D90}},
  \bibinfo{pages}{015032} (\bibinfo{year}{2014}), \eprint{1405.5204}.

\bibitem[{\citenamefont{Gherghetta et~al.}(2015)\citenamefont{Gherghetta, von
  Harling, Medina, Schmidt, and Trott}}]{Cascade7}
\bibinfo{author}{\bibfnamefont{T.}~\bibnamefont{Gherghetta}},
  \bibinfo{author}{\bibfnamefont{B.}~\bibnamefont{von Harling}},
  \bibinfo{author}{\bibfnamefont{A.~D.} \bibnamefont{Medina}},
  \bibinfo{author}{\bibfnamefont{M.~A.} \bibnamefont{Schmidt}},
  \bibnamefont{and} \bibinfo{author}{\bibfnamefont{T.}~\bibnamefont{Trott}},
  \bibinfo{journal}{Phys. Rev.} \textbf{\bibinfo{volume}{D91}},
  \bibinfo{pages}{105004} (\bibinfo{year}{2015}), \eprint{1502.07173}.

\bibitem[{\citenamefont{Cline et~al.}(2015)\citenamefont{Cline, Dupuis, Liu,
  and Xue}}]{Cascade8}
\bibinfo{author}{\bibfnamefont{J.~M.} \bibnamefont{Cline}},
  \bibinfo{author}{\bibfnamefont{G.}~\bibnamefont{Dupuis}},
  \bibinfo{author}{\bibfnamefont{Z.}~\bibnamefont{Liu}}, \bibnamefont{and}
  \bibinfo{author}{\bibfnamefont{W.}~\bibnamefont{Xue}},
  \bibinfo{journal}{Phys. Rev.} \textbf{\bibinfo{volume}{D91}},
  \bibinfo{pages}{115010} (\bibinfo{year}{2015}), \eprint{1503.08213}.

\bibitem[{\citenamefont{Ko and Tang}(2015)}]{Cascade9}
\bibinfo{author}{\bibfnamefont{P.}~\bibnamefont{Ko}} \bibnamefont{and}
  \bibinfo{author}{\bibfnamefont{Y.}~\bibnamefont{Tang}}
  (\bibinfo{year}{2015}), \eprint{1504.03908}.

\bibitem[{\citenamefont{Elor et~al.}(2015)\citenamefont{Elor, Rodd, and
  Slatyer}}]{CascadeAppend1}
\bibinfo{author}{\bibfnamefont{G.}~\bibnamefont{Elor}},
  \bibinfo{author}{\bibfnamefont{N.~L.} \bibnamefont{Rodd}}, \bibnamefont{and}
  \bibinfo{author}{\bibfnamefont{T.~R.} \bibnamefont{Slatyer}},
  \bibinfo{journal}{Phys. Rev.} \textbf{\bibinfo{volume}{D91}},
  \bibinfo{pages}{103531} (\bibinfo{year}{2015}), \eprint{1503.01773}.

\bibitem[{\citenamefont{Heikinheimo and Spethmann}(2014)}]{CascadeRequired1}
\bibinfo{author}{\bibfnamefont{M.}~\bibnamefont{Heikinheimo}} \bibnamefont{and}
  \bibinfo{author}{\bibfnamefont{C.}~\bibnamefont{Spethmann}},
  \bibinfo{journal}{JHEP} \textbf{\bibinfo{volume}{12}}, \bibinfo{pages}{084}
  (\bibinfo{year}{2014}), \eprint{1410.4842}.

\bibitem[{\citenamefont{Boehm et~al.}(2014)\citenamefont{Boehm, Dolan, and
  McCabe}}]{CascadeRequired2}
\bibinfo{author}{\bibfnamefont{C.}~\bibnamefont{Boehm}},
  \bibinfo{author}{\bibfnamefont{M.~J.} \bibnamefont{Dolan}}, \bibnamefont{and}
  \bibinfo{author}{\bibfnamefont{C.}~\bibnamefont{McCabe}},
  \bibinfo{journal}{Phys. Rev.} \textbf{\bibinfo{volume}{D90}},
  \bibinfo{pages}{023531} (\bibinfo{year}{2014}), \eprint{1404.4977}.

\bibitem[{\citenamefont{Cerdeno and Seto}(2009)}]{NMSSMN1}
\bibinfo{author}{\bibfnamefont{D.~G.} \bibnamefont{Cerdeno}} \bibnamefont{and}
  \bibinfo{author}{\bibfnamefont{O.}~\bibnamefont{Seto}},
  \bibinfo{journal}{JCAP} \textbf{\bibinfo{volume}{0908}}, \bibinfo{pages}{032}
  (\bibinfo{year}{2009}), \eprint{0903.4677}.

\bibitem[{\citenamefont{Deppisch and Pilaftsis}(2008)}]{NMSSMN2}
\bibinfo{author}{\bibfnamefont{F.}~\bibnamefont{Deppisch}} \bibnamefont{and}
  \bibinfo{author}{\bibfnamefont{A.}~\bibnamefont{Pilaftsis}},
  \bibinfo{journal}{JHEP} \textbf{\bibinfo{volume}{10}}, \bibinfo{pages}{080}
  (\bibinfo{year}{2008}), \eprint{0808.0490}.

\bibitem[{\citenamefont{Iqbal and Lei}(2015)}]{NMSSMN3}
\bibinfo{author}{\bibfnamefont{S.~T.} \bibnamefont{Iqbal}} \bibnamefont{and}
  \bibinfo{author}{\bibfnamefont{Z.}~\bibnamefont{Lei}}, \bibinfo{journal}{J.
  Phys.} \textbf{\bibinfo{volume}{G42}}, \bibinfo{pages}{095003}
  (\bibinfo{year}{2015}).

\bibitem[{\citenamefont{Wang et~al.}(2013)\citenamefont{Wang, Yang, and
  You}}]{LotsofMistakes}
\bibinfo{author}{\bibfnamefont{W.}~\bibnamefont{Wang}},
  \bibinfo{author}{\bibfnamefont{J.~M.} \bibnamefont{Yang}}, \bibnamefont{and}
  \bibinfo{author}{\bibfnamefont{L.~L.} \bibnamefont{You}},
  \bibinfo{journal}{JHEP} \textbf{\bibinfo{volume}{07}}, \bibinfo{pages}{158}
  (\bibinfo{year}{2013}), \eprint{1303.6465}.

\bibitem[{\citenamefont{Kang et~al.}(2011)\citenamefont{Kang, Li, Li, Liu, and
  Yang}}]{NMSSMN4}
\bibinfo{author}{\bibfnamefont{Z.}~\bibnamefont{Kang}},
  \bibinfo{author}{\bibfnamefont{J.}~\bibnamefont{Li}},
  \bibinfo{author}{\bibfnamefont{T.}~\bibnamefont{Li}},
  \bibinfo{author}{\bibfnamefont{T.}~\bibnamefont{Liu}}, \bibnamefont{and}
  \bibinfo{author}{\bibfnamefont{J.}~\bibnamefont{Yang}}
  (\bibinfo{year}{2011}), \eprint{1102.5644}.

\bibitem[{\citenamefont{Cerdeno et~al.}(2011)\citenamefont{Cerdeno, Huh, Peiro,
  and Seto}}]{NMSSMN5}
\bibinfo{author}{\bibfnamefont{D.~G.} \bibnamefont{Cerdeno}},
  \bibinfo{author}{\bibfnamefont{J.-H.} \bibnamefont{Huh}},
  \bibinfo{author}{\bibfnamefont{M.}~\bibnamefont{Peiro}}, \bibnamefont{and}
  \bibinfo{author}{\bibfnamefont{O.}~\bibnamefont{Seto}},
  \bibinfo{journal}{JCAP} \textbf{\bibinfo{volume}{1111}}, \bibinfo{pages}{027}
  (\bibinfo{year}{2011}), \eprint{1108.0978}.

\bibitem[{\citenamefont{Tang}(2014)}]{NMSSMNMy}
\bibinfo{author}{\bibfnamefont{Y.-L.} \bibnamefont{Tang}},
  \bibinfo{journal}{Nucl. Phys.} \textbf{\bibinfo{volume}{B890}},
  \bibinfo{pages}{263} (\bibinfo{year}{2014}), \eprint{1411.1892}.

\bibitem[{\citenamefont{Allahverdi
  et~al.}(2009{\natexlab{a}})\citenamefont{Allahverdi, Bornhauser, Dutta, and
  Richardson-McDaniel}}]{BML1}
\bibinfo{author}{\bibfnamefont{R.}~\bibnamefont{Allahverdi}},
  \bibinfo{author}{\bibfnamefont{S.}~\bibnamefont{Bornhauser}},
  \bibinfo{author}{\bibfnamefont{B.}~\bibnamefont{Dutta}}, \bibnamefont{and}
  \bibinfo{author}{\bibfnamefont{K.}~\bibnamefont{Richardson-McDaniel}},
  \bibinfo{journal}{Phys. Rev.} \textbf{\bibinfo{volume}{D80}},
  \bibinfo{pages}{055026} (\bibinfo{year}{2009}{\natexlab{a}}),
  \eprint{0907.1486}.

\bibitem[{\citenamefont{Allahverdi et~al.}(2012)\citenamefont{Allahverdi,
  Campbell, and Dutta}}]{BML2}
\bibinfo{author}{\bibfnamefont{R.}~\bibnamefont{Allahverdi}},
  \bibinfo{author}{\bibfnamefont{S.}~\bibnamefont{Campbell}}, \bibnamefont{and}
  \bibinfo{author}{\bibfnamefont{B.}~\bibnamefont{Dutta}},
  \bibinfo{journal}{Phys. Rev.} \textbf{\bibinfo{volume}{D85}},
  \bibinfo{pages}{035004} (\bibinfo{year}{2012}), \eprint{1110.6660}.

\bibitem[{\citenamefont{Allahverdi
  et~al.}(2009{\natexlab{b}})\citenamefont{Allahverdi, Dutta,
  Richardson-McDaniel, and Santoso}}]{BML3}
\bibinfo{author}{\bibfnamefont{R.}~\bibnamefont{Allahverdi}},
  \bibinfo{author}{\bibfnamefont{B.}~\bibnamefont{Dutta}},
  \bibinfo{author}{\bibfnamefont{K.}~\bibnamefont{Richardson-McDaniel}},
  \bibnamefont{and} \bibinfo{author}{\bibfnamefont{Y.}~\bibnamefont{Santoso}},
  \bibinfo{journal}{Phys. Lett.} \textbf{\bibinfo{volume}{B677}},
  \bibinfo{pages}{172} (\bibinfo{year}{2009}{\natexlab{b}}),
  \eprint{0902.3463}.

\bibitem[{\citenamefont{Allahverdi et~al.}(2014)\citenamefont{Allahverdi,
  Campbell, Dutta, and Gao}}]{BML4}
\bibinfo{author}{\bibfnamefont{R.}~\bibnamefont{Allahverdi}},
  \bibinfo{author}{\bibfnamefont{S.~S.} \bibnamefont{Campbell}},
  \bibinfo{author}{\bibfnamefont{B.}~\bibnamefont{Dutta}}, \bibnamefont{and}
  \bibinfo{author}{\bibfnamefont{Y.}~\bibnamefont{Gao}},
  \bibinfo{journal}{Phys. Rev.} \textbf{\bibinfo{volume}{D90}},
  \bibinfo{pages}{073002} (\bibinfo{year}{2014}), \eprint{1405.6253}.

\bibitem[{\citenamefont{Minkowski}(1977)}]{SeeSaw1}
\bibinfo{author}{\bibfnamefont{P.}~\bibnamefont{Minkowski}},
  \bibinfo{journal}{Phys. Lett.} \textbf{\bibinfo{volume}{B67}},
  \bibinfo{pages}{421} (\bibinfo{year}{1977}).

\bibitem[{\citenamefont{Yanagida}(1979)}]{SeeSaw2}
\bibinfo{author}{\bibfnamefont{T.}~\bibnamefont{Yanagida}},
  \bibinfo{journal}{in Proc. of the Workshop on Unified Theory and Baryon
  Number of the Universe (KEK, Tsukuba)} p.~\bibinfo{pages}{95}
  (\bibinfo{year}{1979}).

\bibitem[{\citenamefont{M.~Gell-Mann and Slansky}(1979)}]{SeeSaw3}
\bibinfo{author}{\bibfnamefont{P.~R.} \bibnamefont{M.~Gell-Mann}}
  \bibnamefont{and} \bibinfo{author}{\bibfnamefont{R.}~\bibnamefont{Slansky}},
  \bibinfo{journal}{in Sanibel talk, CALT-68-709 (Feb. 1979), and in
  Supergravity (North Holland, Amsterdam, 1979), p315}  (\bibinfo{year}{1979}).

\bibitem[{\citenamefont{Glashow}(1980)}]{SeeSaw4}
\bibinfo{author}{\bibfnamefont{S.}~\bibnamefont{Glashow}}, \bibinfo{journal}{in
  Quarks and Leptons (Plenum, New York), p. 707}  (\bibinfo{year}{1980}).

\bibitem[{\citenamefont{Mohapatra and Senjanovic}(1980)}]{SeeSaw5}
\bibinfo{author}{\bibfnamefont{R.~N.} \bibnamefont{Mohapatra}}
  \bibnamefont{and}
  \bibinfo{author}{\bibfnamefont{G.}~\bibnamefont{Senjanovic}},
  \bibinfo{journal}{Phys. Rev. Lett.} \textbf{\bibinfo{volume}{44}},
  \bibinfo{pages}{912} (\bibinfo{year}{1980}).

\bibitem[{\citenamefont{Bélanger et~al.}(2015)\citenamefont{Bélanger,
  Boudjema, Pukhov, and Semenov}}]{micrOMEGAs}
\bibinfo{author}{\bibfnamefont{G.}~\bibnamefont{Bélanger}},
  \bibinfo{author}{\bibfnamefont{F.}~\bibnamefont{Boudjema}},
  \bibinfo{author}{\bibfnamefont{A.}~\bibnamefont{Pukhov}}, \bibnamefont{and}
  \bibinfo{author}{\bibfnamefont{A.}~\bibnamefont{Semenov}},
  \bibinfo{journal}{Comput. Phys. Commun.} \textbf{\bibinfo{volume}{192}},
  \bibinfo{pages}{322} (\bibinfo{year}{2015}), \eprint{1407.6129}.

\bibitem[{\citenamefont{Alwall et~al.}(2014)\citenamefont{Alwall, Frederix,
  Frixione, Hirschi, Maltoni, Mattelaer, Shao, Stelzer, Torrielli, and
  Zaro}}]{MadGraph}
\bibinfo{author}{\bibfnamefont{J.}~\bibnamefont{Alwall}},
  \bibinfo{author}{\bibfnamefont{R.}~\bibnamefont{Frederix}},
  \bibinfo{author}{\bibfnamefont{S.}~\bibnamefont{Frixione}},
  \bibinfo{author}{\bibfnamefont{V.}~\bibnamefont{Hirschi}},
  \bibinfo{author}{\bibfnamefont{F.}~\bibnamefont{Maltoni}},
  \bibinfo{author}{\bibfnamefont{O.}~\bibnamefont{Mattelaer}},
  \bibinfo{author}{\bibfnamefont{H.~S.} \bibnamefont{Shao}},
  \bibinfo{author}{\bibfnamefont{T.}~\bibnamefont{Stelzer}},
  \bibinfo{author}{\bibfnamefont{P.}~\bibnamefont{Torrielli}},
  \bibnamefont{and} \bibinfo{author}{\bibfnamefont{M.}~\bibnamefont{Zaro}},
  \bibinfo{journal}{JHEP} \textbf{\bibinfo{volume}{07}}, \bibinfo{pages}{079}
  (\bibinfo{year}{2014}), \eprint{1405.0301}.

\bibitem[{\citenamefont{Sjöstrand et~al.}(2015)\citenamefont{Sjöstrand, Ask,
  Christiansen, Corke, Desai, Ilten, Mrenna, Prestel, Rasmussen, and
  Skands}}]{PYTHIA82}
\bibinfo{author}{\bibfnamefont{T.}~\bibnamefont{Sjöstrand}},
  \bibinfo{author}{\bibfnamefont{S.}~\bibnamefont{Ask}},
  \bibinfo{author}{\bibfnamefont{J.~R.} \bibnamefont{Christiansen}},
  \bibinfo{author}{\bibfnamefont{R.}~\bibnamefont{Corke}},
  \bibinfo{author}{\bibfnamefont{N.}~\bibnamefont{Desai}},
  \bibinfo{author}{\bibfnamefont{P.}~\bibnamefont{Ilten}},
  \bibinfo{author}{\bibfnamefont{S.}~\bibnamefont{Mrenna}},
  \bibinfo{author}{\bibfnamefont{S.}~\bibnamefont{Prestel}},
  \bibinfo{author}{\bibfnamefont{C.~O.} \bibnamefont{Rasmussen}},
  \bibnamefont{and} \bibinfo{author}{\bibfnamefont{P.~Z.}
  \bibnamefont{Skands}}, \bibinfo{journal}{Comput. Phys. Commun.}
  \textbf{\bibinfo{volume}{191}}, \bibinfo{pages}{159} (\bibinfo{year}{2015}),
  \eprint{1410.3012}.

\bibitem[{\citenamefont{Alloul et~al.}(2014)\citenamefont{Alloul, Christensen,
  Degrande, Duhr, and Fuks}}]{FeynRules}
\bibinfo{author}{\bibfnamefont{A.}~\bibnamefont{Alloul}},
  \bibinfo{author}{\bibfnamefont{N.~D.} \bibnamefont{Christensen}},
  \bibinfo{author}{\bibfnamefont{C.}~\bibnamefont{Degrande}},
  \bibinfo{author}{\bibfnamefont{C.}~\bibnamefont{Duhr}}, \bibnamefont{and}
  \bibinfo{author}{\bibfnamefont{B.}~\bibnamefont{Fuks}},
  \bibinfo{journal}{Comput. Phys. Commun.} \textbf{\bibinfo{volume}{185}},
  \bibinfo{pages}{2250} (\bibinfo{year}{2014}), \eprint{1310.1921}.

\bibitem[{\citenamefont{et~al. (Particle Data~Group)}(2014)}]{PDG}
\bibinfo{author}{\bibfnamefont{K.~O.} \bibnamefont{et~al. (Particle
  Data~Group)}}, \bibinfo{journal}{Chin. Phys.} \textbf{\bibinfo{volume}{C38}},
  \bibinfo{pages}{090001} (\bibinfo{year}{2014}).

\bibitem[{\citenamefont{Dib and Kim}(2015)}]{RHNCollider1}
\bibinfo{author}{\bibfnamefont{C.~O.} \bibnamefont{Dib}} \bibnamefont{and}
  \bibinfo{author}{\bibfnamefont{C.~S.} \bibnamefont{Kim}},
  \bibinfo{journal}{Phys. Rev.} \textbf{\bibinfo{volume}{D92}},
  \bibinfo{pages}{093009} (\bibinfo{year}{2015}), \eprint{1509.05981}.

\bibitem[{\citenamefont{Aad et~al.}(2015)}]{RHNCollider2}
\bibinfo{author}{\bibfnamefont{G.}~\bibnamefont{Aad}} \bibnamefont{et~al.}
  (\bibinfo{collaboration}{ATLAS}), \bibinfo{journal}{JHEP}
  \textbf{\bibinfo{volume}{07}}, \bibinfo{pages}{162} (\bibinfo{year}{2015}),
  \eprint{1506.06020}.

\bibitem[{\citenamefont{Khachatryan et~al.}(2015)}]{RHNCollider3}
\bibinfo{author}{\bibfnamefont{V.}~\bibnamefont{Khachatryan}}
  \bibnamefont{et~al.} (\bibinfo{collaboration}{CMS}), \bibinfo{journal}{Phys.
  Lett.} \textbf{\bibinfo{volume}{B748}}, \bibinfo{pages}{144}
  (\bibinfo{year}{2015}), \eprint{1501.05566}.

\bibitem[{\citenamefont{Antusch and Fischer}(2015)}]{RHNCollider4}
\bibinfo{author}{\bibfnamefont{S.}~\bibnamefont{Antusch}} \bibnamefont{and}
  \bibinfo{author}{\bibfnamefont{O.}~\bibnamefont{Fischer}},
  \bibinfo{journal}{JHEP} \textbf{\bibinfo{volume}{05}}, \bibinfo{pages}{053}
  (\bibinfo{year}{2015}), \eprint{1502.05915}.

\bibitem[{\citenamefont{Deppisch et~al.}(2015)\citenamefont{Deppisch,
  Bhupal~Dev, and Pilaftsis}}]{RHNCollider5}
\bibinfo{author}{\bibfnamefont{F.~F.} \bibnamefont{Deppisch}},
  \bibinfo{author}{\bibfnamefont{P.~S.} \bibnamefont{Bhupal~Dev}},
  \bibnamefont{and}
  \bibinfo{author}{\bibfnamefont{A.}~\bibnamefont{Pilaftsis}},
  \bibinfo{journal}{New J. Phys.} \textbf{\bibinfo{volume}{17}},
  \bibinfo{pages}{075019} (\bibinfo{year}{2015}), \eprint{1502.06541}.

\bibitem[{\citenamefont{Das et~al.}(2014)\citenamefont{Das, Bhupal~Dev, and
  Okada}}]{RHNCollider6}
\bibinfo{author}{\bibfnamefont{A.}~\bibnamefont{Das}},
  \bibinfo{author}{\bibfnamefont{P.~S.} \bibnamefont{Bhupal~Dev}},
  \bibnamefont{and} \bibinfo{author}{\bibfnamefont{N.}~\bibnamefont{Okada}},
  \bibinfo{journal}{Phys. Lett.} \textbf{\bibinfo{volume}{B735}},
  \bibinfo{pages}{364} (\bibinfo{year}{2014}), \eprint{1405.0177}.

\bibitem[{\citenamefont{Antusch and Fischer}(2014)}]{RequiredRHNLimit1}
\bibinfo{author}{\bibfnamefont{S.}~\bibnamefont{Antusch}} \bibnamefont{and}
  \bibinfo{author}{\bibfnamefont{O.}~\bibnamefont{Fischer}},
  \bibinfo{journal}{JHEP} \textbf{\bibinfo{volume}{10}}, \bibinfo{pages}{94}
  (\bibinfo{year}{2014}), \eprint{1407.6607}.

\bibitem[{\citenamefont{Das and Okada}(2016)}]{RequiredRHNLimit2}
\bibinfo{author}{\bibfnamefont{A.}~\bibnamefont{Das}} \bibnamefont{and}
  \bibinfo{author}{\bibfnamefont{N.}~\bibnamefont{Okada}},
  \bibinfo{journal}{Phys. Rev.} \textbf{\bibinfo{volume}{D93}},
  \bibinfo{pages}{033003} (\bibinfo{year}{2016}), \eprint{1510.04790}.

\bibitem[{\citenamefont{Das and Okada}(2013)}]{RequiredRHNLimit3}
\bibinfo{author}{\bibfnamefont{A.}~\bibnamefont{Das}} \bibnamefont{and}
  \bibinfo{author}{\bibfnamefont{N.}~\bibnamefont{Okada}},
  \bibinfo{journal}{Phys. Rev.} \textbf{\bibinfo{volume}{D88}},
  \bibinfo{pages}{113001} (\bibinfo{year}{2013}), \eprint{1207.3734}.

\bibitem[{\citenamefont{Wyler and Wolfenstein}(1983)}]{Inverse1}
\bibinfo{author}{\bibfnamefont{D.}~\bibnamefont{Wyler}} \bibnamefont{and}
  \bibinfo{author}{\bibfnamefont{L.}~\bibnamefont{Wolfenstein}},
  \bibinfo{journal}{Nucl. Phys.} \textbf{\bibinfo{volume}{B218}},
  \bibinfo{pages}{205} (\bibinfo{year}{1983}).

\bibitem[{\citenamefont{Mohapatra and Valle}(1986)}]{Inverse2}
\bibinfo{author}{\bibfnamefont{R.~N.} \bibnamefont{Mohapatra}}
  \bibnamefont{and} \bibinfo{author}{\bibfnamefont{J.~W.~F.}
  \bibnamefont{Valle}}, \bibinfo{journal}{Phys. Rev.}
  \textbf{\bibinfo{volume}{D34}}, \bibinfo{pages}{1642} (\bibinfo{year}{1986}).

\bibitem[{\citenamefont{Ma}(1987)}]{Inverse3}
\bibinfo{author}{\bibfnamefont{E.}~\bibnamefont{Ma}}, \bibinfo{journal}{Phys.
  Lett.} \textbf{\bibinfo{volume}{B191}}, \bibinfo{pages}{287}
  (\bibinfo{year}{1987}).

\bibitem[{\citenamefont{Mohapatra}(1986)}]{Inverse4}
\bibinfo{author}{\bibfnamefont{R.~N.} \bibnamefont{Mohapatra}},
  \bibinfo{journal}{Phys. Rev. Lett.} \textbf{\bibinfo{volume}{56}},
  \bibinfo{pages}{561} (\bibinfo{year}{1986}).

\bibitem[{\citenamefont{Zhou}(2012)}]{ZhouYeLing}
\bibinfo{author}{\bibfnamefont{Y.-L.} \bibnamefont{Zhou}},
  \bibinfo{journal}{Phys. Rev.} \textbf{\bibinfo{volume}{D86}},
  \bibinfo{pages}{093011} (\bibinfo{year}{2012}), \eprint{1205.2303}.

\bibitem[{\citenamefont{Navarro et~al.}(1996)\citenamefont{Navarro, Frenk, and
  White}}]{NFW}
\bibinfo{author}{\bibfnamefont{J.~F.} \bibnamefont{Navarro}},
  \bibinfo{author}{\bibfnamefont{C.~S.} \bibnamefont{Frenk}}, \bibnamefont{and}
  \bibinfo{author}{\bibfnamefont{S.~D.~M.} \bibnamefont{White}},
  \bibinfo{journal}{Astrophys. J.} \textbf{\bibinfo{volume}{462}},
  \bibinfo{pages}{563} (\bibinfo{year}{1996}), \eprint{astro-ph/9508025}.

\bibitem[{\citenamefont{Tsai et~al.}(2013{\natexlab{b}})\citenamefont{Tsai,
  Yuan, and Huang}}]{Fittingdsph}
\bibinfo{author}{\bibfnamefont{Y.-L.~S.} \bibnamefont{Tsai}},
  \bibinfo{author}{\bibfnamefont{Q.}~\bibnamefont{Yuan}}, \bibnamefont{and}
  \bibinfo{author}{\bibfnamefont{X.}~\bibnamefont{Huang}},
  \bibinfo{journal}{JCAP} \textbf{\bibinfo{volume}{1303}}, \bibinfo{pages}{018}
  (\bibinfo{year}{2013}{\natexlab{b}}), \eprint{1212.3990}.

\bibitem[{\citenamefont{Iocco et~al.}(2011)\citenamefont{Iocco, Pato, Bertone,
  and Jetzer}}]{ProfileUncertainty}
\bibinfo{author}{\bibfnamefont{F.}~\bibnamefont{Iocco}},
  \bibinfo{author}{\bibfnamefont{M.}~\bibnamefont{Pato}},
  \bibinfo{author}{\bibfnamefont{G.}~\bibnamefont{Bertone}}, \bibnamefont{and}
  \bibinfo{author}{\bibfnamefont{P.}~\bibnamefont{Jetzer}},
  \bibinfo{journal}{JCAP} \textbf{\bibinfo{volume}{1111}}, \bibinfo{pages}{029}
  (\bibinfo{year}{2011}), \eprint{1107.5810}.

\end{thebibliography}
\end{document}